\documentclass[showpacs,preprintnumbers,amsmath,amssymb]{revtex4}%

\usepackage{graphicx}
\usepackage{epsfig}
\usepackage{dcolumn}
\usepackage{bm}

\begin{document}


\title{Polarization rotation for light propagating non-parallel
to a magnetic field in QED vacuum and in a dilute electron gas}

\author{H. P\'erez Rojas}\email{hugo@icmf.inf.cu}
 \altaffiliation{Instituto de Cibernetica, Matematica y Fisica, Calle E
309, Vedado, La Habana, Cuba}
\author{E. Rodr\'{\i}guez Querts}%
 \email{elizabeth@icmf.inf.cu}
\affiliation{Instituto de Cibernetica, Matematica y Fisica, Calle E
309, Vedado, La Habana, Cuba
}%

\date{\today}

\begin{abstract}
The rotation of the polarization vector for light propagating
perpendicular to an external constant external magnetic field $B$,
is calculated in quantum vacuum, where it leads to different photon
eigenmodes of the magnetized photon self-energy tensor for
polarizations along and orthogonal to $B$ (Cotton-Mouton effect in
QED vacuum). Its analogies and differences with Faraday effect are
discussed and both phenomena are calculated for a relativistic
electron gas at low densities, by starting from the low energy limit
of the photon self-energy eigenvalues in presence of $B$. In the
Cotton-Mouton case the polarization vector describes an ellipse
whose axes vary periodically from zero to a maximum value. By
assuming an effective electron density of order $10^3$ cm$^{-3}$ the
quantum relativistic eigenvalues lead to a rotation of the
polarization plane compatible with some of the  limit values
reported by PVLAS experiments. Other consequences, which are
interesting for astrophysics, are also discussed.
\end{abstract}

\maketitle

\section{Introduction}
For light propagating in magnetized quantum vacuum, in a direction
perpendicular to the magnetic field $B$, and polarized along a plane
forming a nonzero angle $\beta$ with $B$, the polarization vector
rotates describing a curve which is a sort of ellipse whose axes
oscillate in time, making the polarization plane to rotate. This
effect is due to the birefringence of magnetized quantum vacuum (for
a medium it is known as Cotton-Mouton effect), which exhibits
different dispersion laws for light polarized along and
perpendicular to $B$. In the present letter we obtain the frequency
of the ellipse axes oscillations by starting from the difference of
the photon self-energy tensor eigenvalues, which we calculate by
starting from the low frequency limit of the expressions obtained by
Shabad \cite{ShabadAP} for the photon self-energy. We find a similar
phenomenon for a photon propagating in a relativistic electron gas,
and from the low density limit of the corresponding eigenvalues
\cite{PerezRojasShabad} we obtain the oscillation frequency of the
ellipse axes, leading to the Cotton-Mouton effect. By a Lorentz
boost parallel to $B$ it results that the rotation occurs for any
photon non-parallel propagation direction, whose polarization is
also non-parallel to $B$.

 We will discuss  in the present paper both cases
 of propagation in quantum vacuum and in a dilute
electron gas, which may arise from the ionization of neutral atoms
or molecules. But, based on general arguments, even in the case that
such background charge is negligible, a dilute neutral gas (for
instance, that remaining in the ultra-high vacuum achieved in the
laboratories, or in astrophysics, the components of some gaseous
nebulae) would exhibit the birefringence effect, if atoms and
molecules are under the action of an axially symmetric  external
field $B$. Their interactions with photons would differ for
polarizations parallel and perpendicular to $B$. As a result, it is
expected to have different frequencies for these two directions,
leading to a Cotton-Mouton rotation. This effect  bears some analogy
to the Faraday rotation, due also to birefringence, which occurs in
a magnetized medium non-invariant under charge conjugation for
photon propagation parallel to $B$. Thus, in QED vacuum Faraday
effect does not exist, but it occurs  if a charge background is
present. For a charged relativistic electron -positron gas  its
dispersion modes and polarizations were found in
\cite{PerezRojasShabad}, and some additional features are discussed
below in Section III. However, the Cotton-Mouton rotation occurs in
vacuum as well as in any other media with no restriction about the
charge conjugation property.   Our results are interesting both in
connection to PVLAS experiments and in astroparticle physics.

 In magnetized QED vacuum the spatial symmetry is explicitly broken by
the field $\textbf{B}$. Electrons and positrons (observable and
virtual) move in bound states characterized by discrete Landau
quantum numbers $0,1,2..$ on the plane orthogonal to $\textbf{B}$
(the quantum version of the classical circular motion), and move
freely along $\textbf{B}$. By using the appropriate vectors
characterizing the reduced symmetry properties and from gauge
invariance,(see \cite{batalin}, \cite{ShabadAP}) one can write the
general tensor structure of the photon self-energy or polarization
operator $\Pi_{\mu \nu}$, which is different from that of isotropic
vacuum $(\mathbf{B}=0)$. The polarization operator can be written as
a linear combination of three basic tensors, which are even
functions of the electromagnetic field tensor ${F}_{\mu \nu}$.

\section{Low energy photon eigenmodes and polarization rotation in vacuum}
To understand what follows, we recall briefly some properties of the
photon eigenmodes in a magnetic field $B$ \cite{ShabadAP}. The
diagonalization of the photon self-energy tensor leads to the
equations $\Pi_{\mu \nu}C^{(i)}_{\nu}=\kappa^{(i)}C^{(i)}_{\mu}$
having three non vanishing eigenvalues
 and three eigenvectors for $i=1,2,3$ (One additional eigenvector is
the photon four momentum vector $k_{\nu}$ whose eigenvalue is
$\kappa^{(4)}=0$). The first three eigenvectors satisfy the four
dimensional transversality condition $C^{(1,2,3)}_{\mu}k_{\mu}=0$),
and are $C^{1}_\mu= k^2 F^2_{\mu \lambda}k^\lambda-k_\mu (kF^2 k)$,
$C^{2}_\mu=F^{*}_{\mu \lambda}k^\lambda$, $C^{3}_\mu=F_{\mu
\lambda}k^\lambda$ (Here $F_{\mu \nu}=\partial_\mu
A_\nu-\partial_\nu A_\mu$ is the electromagnetic field tensor and
$F^*_{\mu \nu}=\frac{i}{2}\epsilon_{\mu \nu \rho \kappa}F^{\rho
\kappa}$ its dual.). By considering $a^{(i)}_\mu (x)$  as the
electromagnetic four vector describing these eigenmodes, its
electric and magnetic fields are ${\bf e^{(i)}}=-\frac{\partial
}{\partial x_0}\vec{a}^{(i)}-\frac{\partial }{\partial {\bf
x}}a^{(i)}_0$, ${\bf h}^{(i)}=\nabla\times\vec{a}^{(i)}$ .  The
polarization vectors are given in detail in  \cite{ShabadJETP},
\cite{ShabadAP}.

 The eigenvalues and eigenvectors
characterize the eigenmodes of wave propagation in the magnetized
vacuum. In what follows we consider sometimes the eigenvectors
normalized to unity (which is valid whenever the magnetic field
$\textbf{B}$ is not taken as zero), and denote them by
$a^{(i)}_\mu$. We must recall here that for propagation
perpendicular to the field, the second mode has polarizations
proportional to
$\textbf{e}^{(2)}=\textbf{b}_{\parallel}(k_{\parallel}^2-\omega^2),
\textbf{h}^{(2)}=-[{\bf b}_{\perp}\times {\bf b}_{\parallel}]$. We
use the set of unit vectors $\textbf{b}_{1,2}$ , (or in general
$\textbf{b}_{\perp}$) orthogonal to $\textbf{B}$, and
$\textbf{b}_{3}=\textbf{b}_{\parallel}$ along it, whereas the third
mode has polarizations $\textbf{e}^{(3)}=-[{\bf b}_{\perp}\times
{\bf b}_{\parallel} ] \omega, \textbf{h}^{(3)}_{\perp}=-{\bf
b}_{\perp}k_{\parallel}$. The first mode is purely longitudinal
${\bf e}^{(1)}=-\textbf{b}_{\perp} \omega$  and non-physical in
quantum vacuum. In a charged medium it is responsible of
longitudinal waves. For propagation along $B$, the modes 1,3 are
transverse with polarizations proportional to ${\bf
e}^{(1)}=-\textbf{b}_{\perp} \omega$ and $\textbf{e}^{(3)}=-[{\bf
b}_{\perp}\times {\bf b}_{\parallel} ] \omega$, respectively,
whereas the second is pure longitudinal
$\textbf{e}^{(2)}_{\parallel}=\textbf{b}_{\parallel}(k_{\parallel}^2-\omega^2)$
and consequence it has meaning only in a charged medium. The vectors
${\bf k}_{\perp}$ and ${\bf k}_{\parallel}$ are the components of
$\textbf{k}$ across and along $\bf B$. The previous formulae refer
to the reference frame which is at rest or moving parallel to $B$.

These results indicate that the dispersion equations and
polarizations of light in a magnetic field are different along
different directions in space. Let us consider  a plane polarized
wave $\textbf{E}=\textbf{f} e^{i\omega t}$ propagating perpendicular
to $B$ and whose polarization vector $\textbf{f}$ is in a direction
forming a nonzero angle $\beta$ with $B$. It can be decomposed in
two waves having polarizations $\textbf{f}_2=\textbf{f} cos \beta$,
$\textbf{f}_3=\textbf{f} sin \beta$ along and perpendicular to
$B$.\textit{Thus, after entering in the magnetized quantum vacuum,
each one of these components obey the dispersion law dictated by the
corresponding eigenmode, and propagate with different polarization
and frequency; birefringence is produced}. The frequencies are
$\omega_2=\omega +\Delta \omega_2$ and $\omega_3=\omega + \Delta
\omega_3$, where $\Delta \omega_{2,3}$ are the contributions from
the corresponding eigenvalues of the polarization operator
$\kappa_{2,3}$ respectively, obtained after solving the dispersion
equations $k^2-\kappa^{(i)}=0$. At a point of space we can write the
projections of the electromagnetic wave $\textbf{E}$ as $E_2= f_2
cos (\omega +\Delta \omega_2)t$ and $E_3= f_3 cos (\omega +\Delta
\omega_3)t$. The polarization vector starts to rotate, since we have
here the well known problem of the resultant curve from orthogonal
oscillations whose frequencies differ in a small fraction of the
main frequency.

We will give below a quantitative estimate of the rotation
frequency. In the low frequency limit, under the additional
condition $k_{\perp}^2<<2eB\hbar c$, the scalars $\kappa^{(2)}$ and
$\kappa^{(3)}$ have been approximated (see Appendix) by expanding
the general expression of the photon self-energy tensor obtained in
\cite{ShabadAP} and taking the first order in $B^2$. The expansion
is made also in terms of the relativistic invariant variables
$z_1=k_{\parallel}^2-\omega^2$, $k_{\perp}^2$, respectively as
(notice that for propagation perpendicular to $B$ we have
$z_1=-\omega^2$.)
\begin{eqnarray}
\kappa^{(2)} &=&\frac{\alpha e^2 B^2 \hbar^4 }{9 \pi m^4 c^6}
\left(\omega^2+\frac{2}{5}k_{\perp}^2 \right),  \label{S_0}\\
\kappa^{(3)} &=& \frac{\alpha e^2 B^2 \hbar^4}{9 \pi m^4 c^6} \left(
-\frac{2 \omega^2}{5}+\frac{6 k_{\perp}^2}{5} \right). \label{T_0}
\end{eqnarray}

These quantities contain the contribution of all Landau levels of
virtual electron-positron pairs. The frequency $\epsilon$ varies in
the present approximation as  $B^2$. On the light cone
$k_{\perp}^2+z_1 =0$, (\ref{S_0}), (\ref{T_0}) agree with earlier
values reported by Dittrich et al \cite{Dittrich:1998gt} in
calculating the vacuum birefringence in a strong magnetic field.

 The estimated ratio of frequencies
$\epsilon/\omega =(\kappa^{(2)}-\kappa^{(3)})/k_{\perp}\omega$ for
$\omega \sim 10^{15}$ rad/s, is $\epsilon/\omega \sim 10^{-21}$.

To find the curve described by the polarization vector we shall write
$\omega'=\omega +
\Delta \omega_2$ for simplicity, and by eliminating $\omega'$ from
the expressions for $E_2, E_3$ one gets

\begin{equation}
f_2^2 E_3^2 -2f_2 f_3 E_2 E_3 cos \epsilon t + f_3^2 E_2^2= f_2^2
f_3^2 sin^2 \epsilon t  \label{lp}
\end{equation}

\begin{figure}[t]
   \epsfig{file=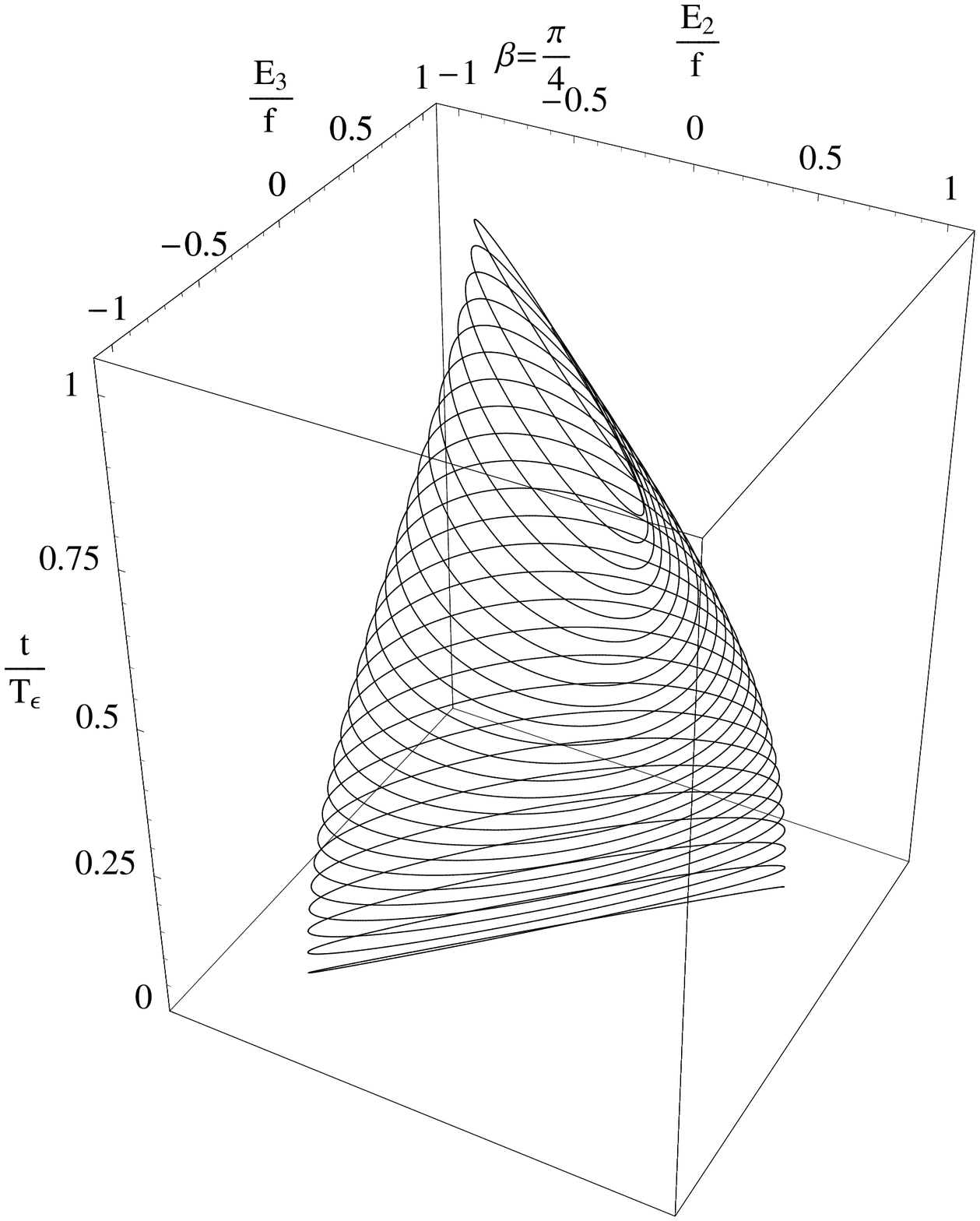,width=13pc}
\epsfig{file=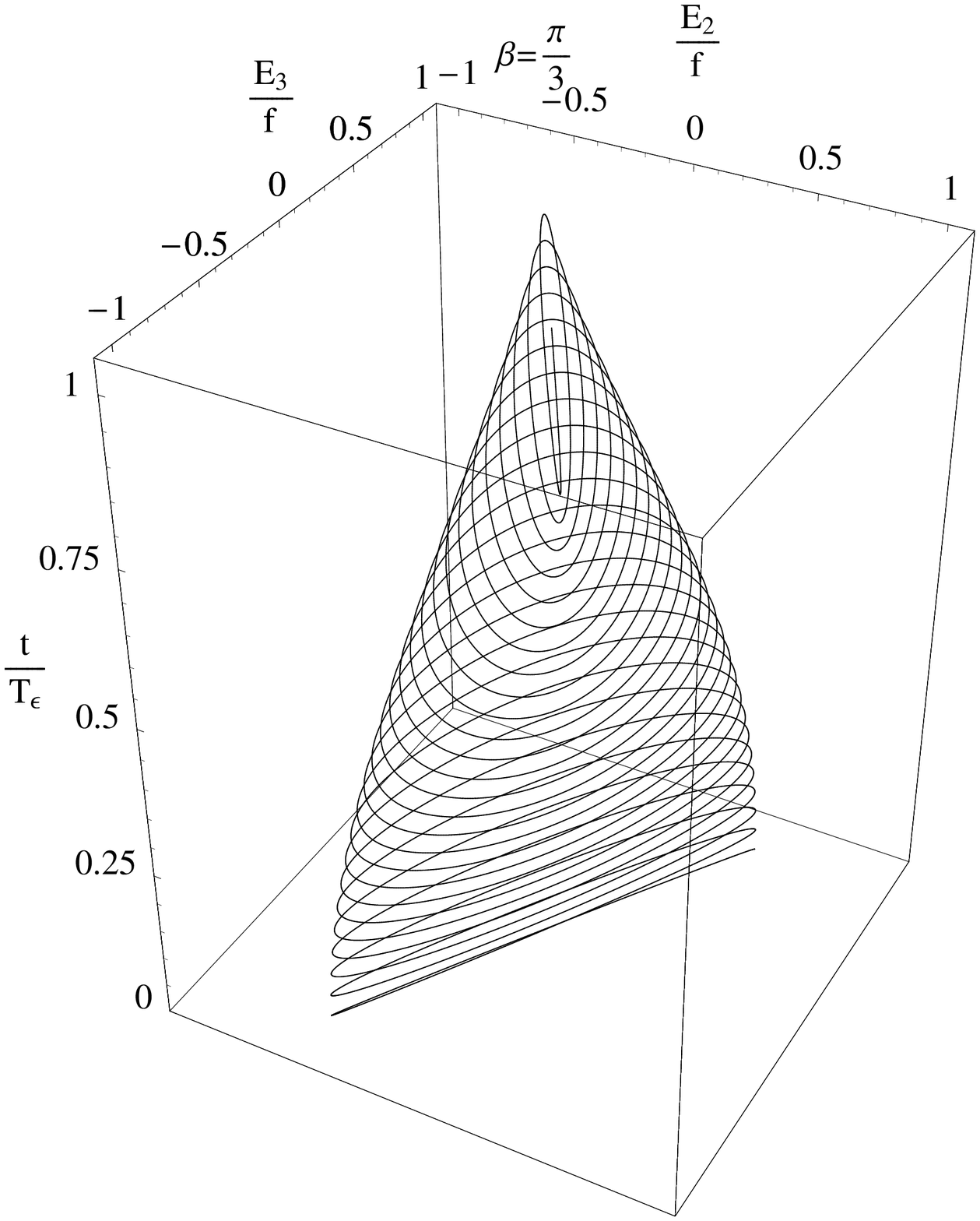,width=13pc}
\epsfig{file=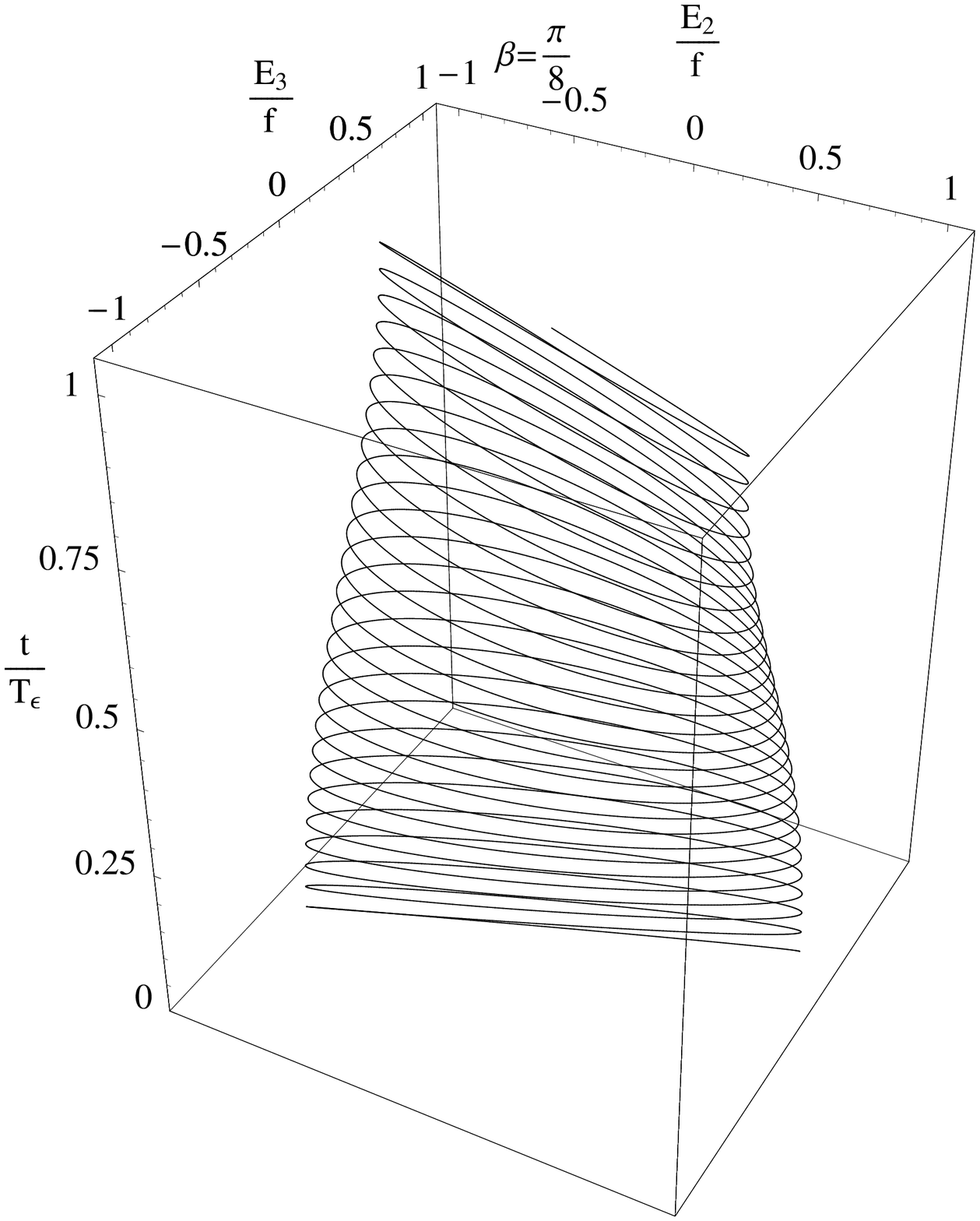,width=13pc}
   \caption{The ellipse described by $E_{2}, E_{3}$ is drawn here for three
    values of $\beta$, along the time axis. We have exaggerated the ratio $\epsilon/\omega$ to $ \sim 10^{-2}$, to make
    the rotation of the polarization plane more evident.\label{lightpol}}
   \end{figure}

The curve representing (\ref{lp}) is  an ellipse with oscillating
axes, expressed in terms of $E_2, E_3$, and $f_2, f_3$. For
$\epsilon t =0, \pi$, it degenerates in two straight lines $E_3= \pm
\tan \beta E_2$. For $\epsilon =\pi/2, 3\pi/2$ we have respectively
two ellipses. The ellipse starts with its minor axis being zero and
its major axis along the initial plane of polarization. The major
axis oscillates and decreases, while the minor axis oscillates
increasing, up to a configuration in which the major axis is zero
and the minor axis is maximum and passes along the other straight
line. Starting from this configuration, the situation is reversed.
(See Fig. \ref{lightpol}). This effect makes the rotation difficult
to be observed in some intervals of time, when one of the semi-axes
is very close to zero. We will see below that if some adequate
density of electrons exists in the medium, the rotation of the
polarization plane is produced at a frequency $\epsilon$ much larger
than in the quantum vacuum, allowing it to be observable at magnetic
fields and frequencies of the same order as those of PVLAS
experiments.

\section{The case of a medium}
 The case of  photons propagating in a relativistic
electron-positron ($e^{\pm}$) medium at temperature $\beta^{-1}\neq
0$  and non-zero particle density (chemical potential $\mu \neq 0$)
is of especial interest \cite{PerezRojasShabad}. The one-loop
diagram describing the process accounting for the photon self-energy
interaction contains, in addition to the virtual creation and
annihilation of the pair, the process of absorption and subsequent
emission of one photon by the electron and/or positron. Concerning the average particle densities,
for $\mu \neq 0$ but $\beta^{-1}=0$, the average
positron density is zero. This is usual in most laboratory
conditions.

Also, even in the case of high vacuum conditions of the experiment
\cite{PVLAS2}, we expect that a density of order $10^8$ molecules
remain. This is far from being pure quantum vacuum, but a medium
which is not invariant under charge conjugation. We assume the
existence of a low density electron gas may  due to ionization of a
fraction of these molecules which might be produced both by the
action of the laser beam and the magnetic field intensity. In such
case, their equilibrium with ions might be described by some
chemical potential $\mu \neq 0$ . The ions also contribute to the
self-energy tensor, but due to their large mass, their contribution
is much smaller than that of the electron gas. In such a medium, an
additional longitudinal electromagnetic wave component is also
possible, saturating the three spatial degrees of freedom. Two other
tensors, odd in the chemical potential $\mu$ and anti-symmetric in
the tensor indices $\rho,\nu$ contribute to $\Pi_{\rho \nu}$ through
new basic scalars named $V, R$. The resulting eigenmodes are, in
general, polarized elliptically, and in some cases are combinations
of transverse and longitudinal waves \cite{PerezRojasShabad}.

For propagation along the external field $B$ we name
$\kappa_{2}=S_m$ and $\kappa_{1,3}=T_m \pm R$, where $S_m, T_m, R$
are  scalars depending on temperature and density as well as on
$k_\mu$ and $B$. The second mode is a pure longitudinal wave, whose
electric polarization vector
$\textbf{e}^{(2)}=(\omega^2-k^2_{\parallel})
\textbf{b}_{\parallel}\neq 0$ since $k_{\perp}=0$ but
$\omega^2-k^2_{\parallel}$ does not necessarily vanish in the
medium. The transverse modes are
\begin{equation}
 C^{\prime 1, 3}_\mu = R(C^{3}_\mu \pm i C^{1}_\mu)\label{circ}
\end{equation}
\noindent and describe circularly polarized waves in the plane
orthogonal to $\textbf{B}$ having different eigenvalues. Up to a
normalizing factor, their  electric polarization vectors can be
written respectively as $\textbf{e}^{(1,3)}=
(\textbf{b}^{(1)}_{1}\pm i \textbf{b}^{(3)}_{2} )$, corresponding to
the eigenvalues $T_m \pm R$, leading to the well-known phenomenon of
Faraday effect. Let us consider two electromagnetic waves whose
polarization vectors are proportional respectively to
$\textbf{e}^{(1,3)}$, ($\textbf{h}^{(1,3)}=
\textbf{k}\times\textbf{e}^{(1,3)}$).

 If we also assume that $T_m, R \ll \omega$, one can write
approximately $\omega_{\mp}=\sqrt{\textbf{k}^2- T_m}\mp \epsilon'$,
where $\epsilon'=R/2\sqrt{\textbf{k}^2-T_m}$. Thus, a superposition
of both modes having equal amplitudes leads to the following wave
\begin{eqnarray}
\textbf{E}&=& A Re[ \textbf{e}^1 e^{i(\textbf{k}\cdot
\textbf{x}-\omega_{-}t)}+ \textbf{e}^3 e^{i(\textbf{k}\cdot
\textbf{x}-\omega_{+} t)}]\\ \nonumber
 &=&A[\textbf{e}^1 e^{-i \epsilon't}+ \textbf{e}^3 e^{ i
\epsilon't}]Re e^{i(\textbf{k}\cdot \textbf{x}-\omega t)}\\
\nonumber &=& A(\textbf{b}^1 \cos \epsilon't+ \textbf{b}^2 \sin
\epsilon't )Re e^{i(\textbf{k}\cdot \textbf{x}-\omega t)} ]
\label{Faraday}
\end{eqnarray}
which shows that the polarization of the propagating wave rotates
counterclockwise with frequency $\epsilon'$, describing a
circumference. We recall that as the system has rotational symmetry
around $\textbf{B}$, the direction of the orthogonal eigenvectors
$\textbf{b}^{1,3}$ may be taken arbitrarily.

Concerning the propagation perpendicular to the field, we have the
eigenvalues $\kappa_{2}=S_m$, $\kappa_{1,3}=[ (P-T_m)\pm
\sqrt{(P-T_m)^2-4 R^2}]/2$ which describe two waves whose
polarization vector rotates in the plane orthogonal to $B$. However,
for very low charge densities, the quantities $P$ and $R$ are
negligibly small as compared with $T_m$ for perpendicular
propagation. This means that the eigenvalues $\kappa_{1,3}$ can be
approximated as $\kappa_{3}$, $0$ respectively, the $3$-rd mode
being plane polarized orthogonal to $B$. Thus, a wave entering in
the magnetized medium propagating orthogonal to $B$, with
polarization vector forming some angle $\pi/2>\beta>0$ with
$\textbf{B}$, has components along the polarizations of modes $2,3$.
This means that the polarization vector \textit{would describe an
ellipse with oscillating axes, like the curve mentioned previously
in the vacuum case}. Below it is shown that its frequency of
rotation is approximately a linear function of the electron density,
$N_e$.

 For comparison with
PVLAS results, we take frequencies $\omega \sim 1,8 \times 10^{15}$
rad/s, and magnetic field $B \sim 5.5 \times 10^4 G$, and we may
simplify the expressions given in the Appendix for $T_m$ and $F$.
Let us call the density of particles by $N_e =\sum_n (eB/\hbar^2
c^2) \int dp_3 n_e$, where $n_e$ is the electron Fermi distribution
function. We must stress here that $n_e$ stands for the low
temperature limit of the sum of electrons plus positrons densities
$n_e + n_p$, ($n_p$ is negligible small at low temperatures), since
the Cotton-Mouton rotation exists in any case, even in a neutral
electron-positron gas, where $n_e=n_p$ as different from the Faraday
rotation, which depends on the net charge $n_e - n_p$. We may
approximate $(\ref{T})$ and $(\ref{S})$ given in the Appendix as
\begin{eqnarray}
T_m &=& -\frac{2\alpha }{\pi} \lambda^3 m^2 c^4 N_e \label{T_m}\\
S_m &=& -\frac{\alpha }{\pi} \lambda^3 m^2 c^4  N_e, \label{S_m}
\end{eqnarray}

since $(\ref{T})$ and $(\ref{S})$  under the conditions of
propagation perpendicular to $B$, $k_{\perp}^2 \ll m^2 c^4$ and near
the light cone ($k_{\perp}\sim \omega$) are approximately
independent of the photon momentum and energy. Here $\lambda$ is the
electron Compton wavelength. The dependence on $B$ is contained in
the term $N_e$ (see below).  We assume that to have charge
neutrality one must add to (\ref{T_m}, \ref{S_m}) a similar term in
which the electron density is replaced by the ion density and the
electron mass by the ion mass. Such term, however, is al least
$10^{-3}$ times smaller than expressions (\ref{T_m}),(\ref{S_m}),
and may be ignored in the calculations.

For the previous values of $B$ and $\omega$, we have finally $\Delta
\omega = (S_m-T_m)/ k_{\perp}\sim\frac{\alpha }{\pi} \lambda^3 m^2
c^4 N_e/k_{\perp} \sim 5,6 \times 10^{-34}N_e$. For $N_e \sim 10^3$,
$\epsilon \sim  10^{-31}erg = 10^{-4}$ rad/s, leading to a period $T
\sim 10^4$ s. The angle rotated in a length of one meter is of order
$10^{-13}$ rad.

Eqs. (\ref{T_m}),(\ref{S_m}) are valid  for any value of the
electron density and energy. Their dependence on the magnetic field
is contained in the density of particles term: for constant chemical
potential and zero temperature, the dependence on $B$ comes from the
degeneracy factor $eB$ and from the energy eigenvalues $E_n$,
especially for nonzero temperature. However, another dependence
might come from the chemical potential $\mu$, which may be
$B$-dependent.

Thus, for having a Cotton-Mouton rotation comparable with the limits imposed by the last
results of PVLAS experiments \cite{PVLASII}, if \textit{ on the
average, and in the process of the experiment, the system behaves as
if having a number of free electrons of order $0.001$ percent of the
remnant molecules}. This is enough to give a figure $10^3$ times the
pure quantum vacuum effect.

The Fermi-Dirac degenerate distribution might not be suitable to
describe $n_e$ in some cases, for instance, at nonzero temperature
and very low densities. Thus, although the experimental PVLAS
conditions are made at temperatures near zero, the observed
phenomenon occurs far from equilibrium. For a fixed magnetic field
we may assume that on the average it behaves equivalently to an
equilibrium system having a constant chemical potential and an
effective temperature of several Kelvin degrees, and use the
Boltzmann distribution $n_e= \sum \alpha_n \int dp_3 e^{-((E
-\mu))\beta}$ ($\alpha_n =2-\delta_{0n}$), as an approximation. For
$E$ we take the non-relativistic limit by expanding it as $E=m
c^2+p^2/2m + eB\hbar n/m c$. We call $A = e^{-\mu'\beta}$, where
$\mu'=m c^2-\mu$, and gets easily
\begin{equation}
N_e = A  \frac{e B }{\lambda_D \hbar c}\coth \frac{e B \hbar
\beta}{m c} \label{net}
\end{equation}
where $\lambda_D=\hbar/\sqrt{2\pi m/\beta}$ is the De Broglie
thermal wavelength. For for $\mu'$ constant and $X<1$, by taking
$coth X \sim (1/X + X/3)$, we conclude that $N_e$, and in
consequence $\epsilon$ grows with $B$ as $a + bB^2$, where
$a=m/\lambda_D \hbar^2$,$b=e \hbar \beta/mc$ are temperature
dependent parameters. For $T$  of order of few Kelvin degrees, the
dominant term is $a$ which is independent of $B$. But  the chemical
potential might be dependent on $B$, which is more realistic for
 the ionization process. Thus, for the dilute gas under the
influences of both the laser and magnetic field it is  expected for
$T_m$, $S_m$ a dependence on $B$ more complex than in the quantum
vacuum case (i.e. not proportional to $B^2$)

We must observe that formulae (\ref{T},\ref{S}) in the Appendix may
be applied even in the case of the so-called \textit{hot vacuum}, a
system of electron-positron pairs in equilibrium with photons at
high very high temperature ($\beta^{-1}\geq m c^2\sim 10^9$
${}^\circ$K), Such a system have zero net charge, and is invariant
under charge conjugation. Such system (which might exist in neutron
stars magnetospheres) does not exhibit the Faraday rotation but it
shows the Cotton-Mouton rotation of the polarization plane. Even
more, as hot vacuum implies $\mu = 0$, and the Boltzmann
distribution is a good approximation for it \cite{Kratkie} one can
use the expressions below (\ref{T_m},\ref{S_m}) and (\ref{net}) by
taking $A=2 e^{-\beta mc^2}$. It is seen from (\ref{net}) that by
increasing temperature, the density of particles, and in
consequence, the frequency induced by the magnetic field, decreases.

In concluding, we want to emphasize that the Cotton-Mouton rotation
of the polarization plane for photons propagating orthogonal to $B$,
although bears some analogy to Faraday effect, differs from it in
two main facts: Faraday rotation is circular (which is related to
the axial symmetry determined by the preferred direction of magnetic
field $B$, whereas the Cotton Mouton rotation is a more complex
curve; Faraday rotation occurs only if the system is not invariant
under charge conjugation, whereas Cotton-Mouton rotation does not
require non-invariance under charge conjugation. Thus, it occurs in
QED vacuum,  in a medium containing an electron-ion system, as well
as in a neutral molecular or electron-positron gas. Even if the
remnant ionized molecules density would be negligible small, the
medium would show birefringence properties, with the consequent
rotation of the polarization plane. The Cotton-Mouton rotation is
expected to act on the polarization plane of radiation propagating
across intergalactic media.

The authors thank the Abdus Salam ICTP support under OEA-ICTP Net-35
 Grant. They thank M. Chaichian, A. Tureanu and A. Zepeda for several comments and discussions,
 to A. Gonzalez, B.Rodriguez and S. Villalba for valuable
 suggestions, and especially to A.E. Shabad for several important remarks.

 \section{Appendix:}

\textbf{a)}We write the explicit expressions for $S$ and $T$ in
quantum field theory \cite{ShabadAP}, in which it is taken
$\hbar=c=1$, and in statistics \cite{PerezRojasShabad}. In the first
case,by calling $z_2\equiv k_{\perp}^2$, the scalars $\kappa_{2,3}
=S, T$ are given by
\begin{equation}
\kappa_{2,3}=\frac{2\alpha}{\pi}\int_0^\infty d\tau \int_1^1
d\eta[\rho_{2,3}C]e^{M- \frac{N}{2\tau}}e^{-\tau m^2} \label{Sh}
\end{equation}
where
\[
 C=\frac{eB}{\sinh eB}, N=\frac{1-\eta^2}{4},  M=-\frac{1}{2}z_2 M'-\frac{1}{2}z_1 N
\]
\[
\eta^{\pm}=\frac{1 \pm
 \eta}{2},  M'=\frac{\sinh eB\tau \eta^{+}\sinh
eB\tau \eta^{-}}{eB\sinh eB\tau}
\]
\[
\rho_2= -\frac{1}{2}z_1 \cosh eB\tau N -\frac{1}{2}z_2 Q
\]
\[
\rho_3=-\frac{1}{2}z_2 C M'-\frac{1}{2}z_1 Q
\]
\[
Q=\eta^{-} \frac{\sinh eB\tau \eta^{+} \cosh eB\tau \eta^{+}}{\sinh
eB\tau}
\]
One can expand (\ref{Sh}) in powers of $z_1$, $z_2$. To first order
 one gets
\[
\kappa_{2,3}(z_1,z_2)=z_1(\frac{\partial \kappa_{2,3}}{\partial
z_1})_{z_1=z_2=0}  + z_2(\frac{\partial \kappa_{2,3}}{\partial
z_2})_{z_1=z_2=0},
\]
since $\kappa_{2,3}(0,0)=0$. By taking the expansion in powers of
$B^2$  also to first order, one gets eqs. (\ref{S_0}), (\ref{T_0}).

\textbf{b)} We shall now write from
\cite{PerezRojasShabad},\cite{PerezRojas} the explicit expressions
for the scalars $R$, $T_m$ and $S_m$ for $\omega^2 \hbar^2 \ll 2eB
\hbar c$, in the case of a charged medium, i.e. , in relativistic
quantum statistics. We have
\begin{equation}
T_m = -\frac{eB\hbar c}{4\pi^2}\sum_{n,n'}F^{(2)}_{n,n'}\int_{
-\infty}^{\infty}\frac{dp_3}{E_q}[1-G](n_e + n_p-1) \label{T}
\end{equation}
where $G=[(H + N)K]/|Q|^2$, where $H=z_1 + 2eB\hbar c(n+n'))$,
$K=z_1 + 2eB\hbar c(n'-n)$, $|Q|^2=K^2-4 \omega^2 E_q^2$, and
$n_e$,$n_p$ are the electron and positron Fermi-Dirac distribution
functions. The quantity  $-1$ subtracted to them stands for the
quantum field limit $\mu \to 0$, $\beta^{-1}=0$, whose contribution
to order $B^2$ was discussed previously. The quantity $E_q=
\sqrt{m^2 c^4 +c^2p_3^2 +2eB\hbar c n}$ is the electron-positron
energy eigenvalue. The quantities $F^{(2)}_{n,n'}$, $N=4
N^{(1)}_{n,n'}$ are defined in \cite{PerezRojasShabad}, as well as
$F^{(1,3)}_{n,n'}$ and $G^{(1)}_{n,n'}$, used below, in terms of the
Laguerre functions of the variable $k_{\perp}^2/2eB$. They have the
property $F^{(2,3)}_{n,n'}(0)= \delta_{n, n'-1} \pm \delta_{n-1,
n}$, $F^{(1)}_{n,n'}(0)= \delta_{n-1, n'-1}+\delta_{n, n'}$,
$G^{(1)}_{n,n'}(0)=2(\delta_{nn'}+\delta_{0n})$,$N^{(1)}_{n,n'}(0)=0$
\cite{PerezRojas}.

Now we give the corresponding expression for $R$. We have
$R=\sqrt{\frac{k^2}{z_1}}F$, where
\begin{equation}
F = -\frac{eB\hbar c}{4\pi^2}\sum_{n,n'}F^{(3)}_{n,n'}\int_{
-\infty}^{\infty}\frac{dp_3}{E_q}[H](n_e - n_p).
\end{equation}
The term $F$ arises due to the electron-positron charge asymmetry.
It vanishes in the  limit $\mu \to 0$, but is nonzero in absence of
positrons, whenever the electron density $n_e \neq 0$.

The expression for $S_m$, which in the present case contributes to
the expansion to first order in $\omega^2$.

\begin{eqnarray} S_m& =& -\frac{eB\hbar
c}{2\pi^2}\sum_{n,n'}\int_{
-\infty}^{\infty}\frac{dp_3}{E_q}[L]\frac{K}{|Q|^2}(n_e + n_p-1)
\label{S}
\end{eqnarray}
where $L=2m^2 c^4 + eB\hbar
c(n+n'))F^{(1)}_{n,n'}+eB(nn')^{1/2}G^{(1)}_{nn'}$. For
$\beta^{-1}\to 0$, (i.e., for all practical purposes in most
laboratories) the positron contribution vanishes in
(\ref{T},\ref{S}), and we omit it in what follows). For propagation
parallel to $B$, $k_{\perp}=0$ and the argument $F^{(1, 2)}_{n,n'}$
vanish. In the approximation assumed above ($\omega^2 \hbar^2 \ll
2eB \hbar c$)these expressions are actually valid also for
propagation perpendicular to $\textbf{B}$. This makes $\int (n_e
dp_3)/E_q$ the dominant term in the integrals in $T_m, S_m$.

\textbf{c)}We return to the Faraday rotation problem. We have
\begin{equation}
F = -\frac{\alpha}{2\pi}\frac{m c^2 \lambda^3 eB\hbar c N_e}{
\hbar\omega}
\end{equation}

For propagation parallel to $B$ we have,
\[
T_m \sim 10^{-45}N_e,\\
R=F \sim  10^{-48}N_e
\]
By dividing  $R$ by the factor $|\textbf{k}|\sim 10^{-12}$ erg we
get
\[
\Delta \omega=R/k \sim  10^{-36} N_e
 \]
 Thus, $\epsilon'/\omega =\Delta \omega/\omega \sim 10^{-24}N_e$. A density of $10^6$ electrons/cm$^3$, would give
  a Faraday rotation frequency whose  order of magnitude is comparable with PVLAS
 rotation.

 .

\end{document}